\documentclass[showpacs,preprintnumbers,amsmath,amssymb,prl,twocolumn,nobibnotes]{revtex4}

\usepackage{placeins}
\usepackage{graphicx}
\usepackage{bm}
\usepackage{dcolumn}
\usepackage{array}
\usepackage{color}

\usepackage{changes}

\begin{document}

\title{Quantum Engineering of a Low-Entropy Gas of Heteronuclear Bosonic
Molecules \\ in an Optical Lattice}
\author{Lukas Reichs\"{o}llner$^{1}$}\thanks{These authors contributed equally
to this work.}
\author{Andreas Schindewolf$^{1}$}\thanks{These authors contributed equally to
this work.}
\author{Tetsu Takekoshi$^{1,2}$}
%\author{Beatrix Mayr$^{1}$}
%\author{Silva Mezinska$^{3}$}
\author{Rudolf Grimm$^{1,2}$}
\author{Hanns-Christoph N\"{a}gerl$^{1}$}
\affiliation{
$^{1}$Institut f\"ur Experimentalphysik, Universit\"at Innsbruck,
6020 Innsbruck, Austria
\\
$^{2}$Institut f\"ur Quantenoptik und Quanteninformation,
\"Osterreichische Akademie der Wissenschaften, 6020 Innsbruck, Austria
}

\date{\today}

\begin{abstract}
We demonstrate a generally applicable technique for mixing two-species quantum
degenerate bosonic samples in the presence of an optical lattice, and we employ
it to produce low-entropy samples of ultracold $^{87}$Rb$^{133}$Cs Feshbach
molecules with a lattice filling fraction exceeding 30\%. Starting from two
spatially separated Bose-Einstein condensates of Rb and Cs atoms, Rb-Cs atom
pairs are efficiently produced by using the superfluid-to-Mott insulator
quantum phase transition twice, first for the Cs sample, then for the Rb
sample, after nulling the Rb-Cs interaction at a Feshbach resonance's zero
crossing. We form molecules out of atom pairs and characterize the mixing
process in terms of sample overlap and mixing speed. The dense and ultracold
sample of more than 5000~RbCs molecules is an ideal starting point for
experiments in the context of quantum many-body physics with long-range dipolar
interactions.
\end{abstract}

%     quantum phase transitions, atoms in optical lattices, ultrcold and
%     trapped gases
% PACS, the Physics and Astronomy Classification Scheme.
\maketitle

Samples of dipolar ground-state molecules with low
entropy offer a platform for exploring new areas of quantum many-body
physics and related fields. Because of their long-range, spatially
anisotropic interaction they have been proposed to enable investigations
into novel forms of quantum matter, e.g., supersolidity, unconventional
manifestations of superfluidity, and novel types of quantum magnetism
\cite{Trefzger2011udg, Baranov2012cmt, Lahaye2009tpo}. They are expected
to allow the realization of many-body spin systems \cite{Hazzard2013ffe}
with, in principle, local spin control and readout. In particular, they
promise the study of dynamical processes in such systems, e.g., on
many-body spin transport and inhibition thereof \cite{Deng2016qlf}. In
addition, with the exquisite control over all quantum degrees of freedom, they
offer the possibility of implementing quantum simulation protocols
\cite{Bloch2012qsw} that require genuine and strong long-range interactions.

%%%%%%%%%%%%%%%%%%%%%%%%%%%%%%%%%
\begin{figure}[tbp]
\includegraphics[width=\columnwidth]{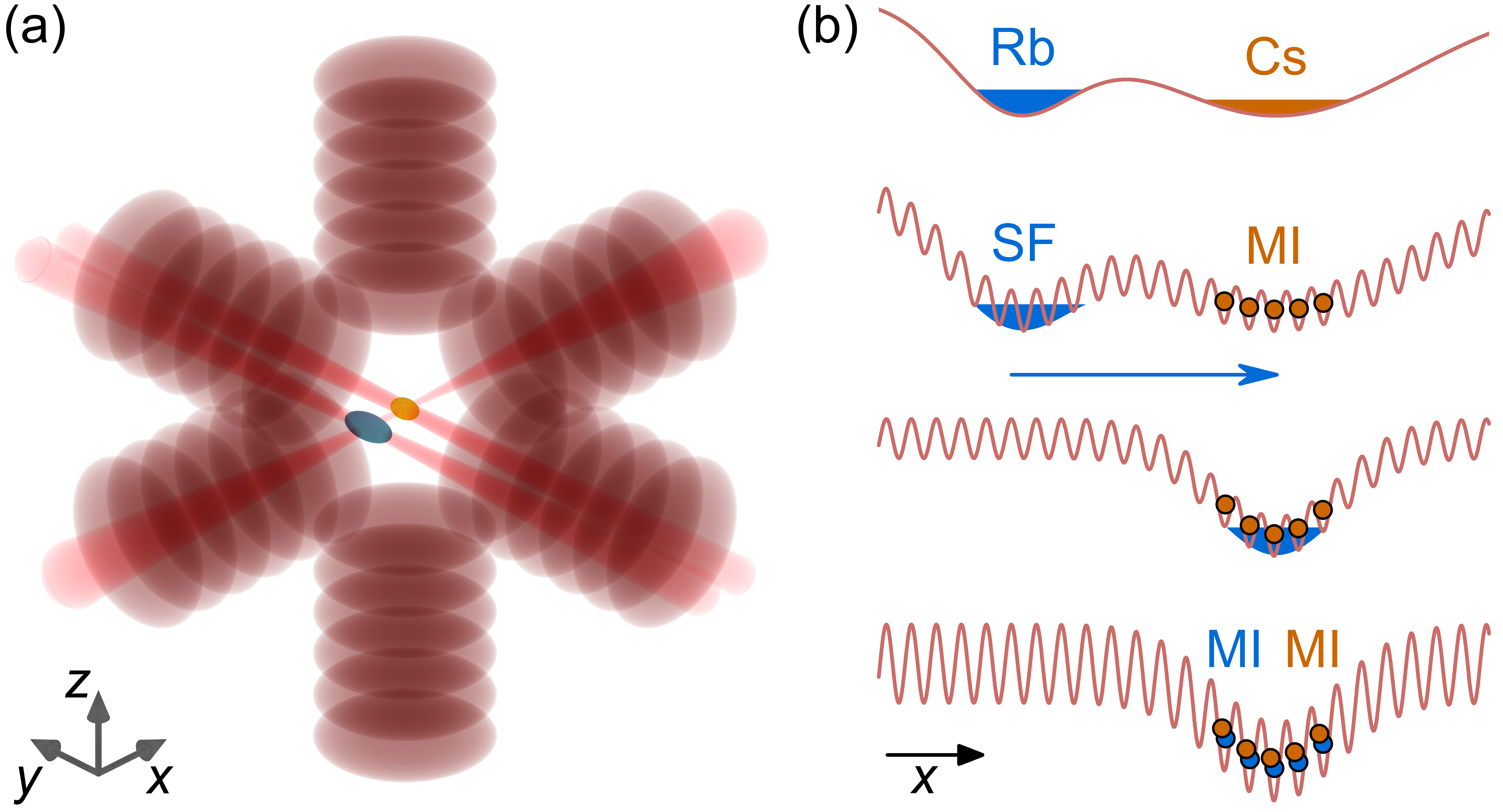}
\caption{Simplified experimental
optical trap setup and overlap strategy.
(a)~Bose-Einstein condensates of Rb (blue) and Cs (orange) are initially
produced in separate crossed dipole traps. They are merged along the
direction of one of the trapping beams ($x$~direction) at the center of
a six-beam optical lattice (standing waves as indicated, center part
omitted for clarity). (b)~Starting from the two BECs, the lattice first
induces a one-atom Mott insulator (MI) for the Cs sample. The Rb sample,
yet superfluid (SF), is brought into overlap with the Cs sample by
precisely moving the underlying dipole trap beam. The lattice is raised
further to create a double-species MI with high Rb-Cs atom pair fraction.}
\label{Fig1}
\end{figure}
%%%%%%%%%%%%%%%%%%%%%%%%%%%%%%%%%

The production of low-entropy samples of rovibronic ground-state molecules is
challenging. To date, the regime of nanokelvin molecular temperatures has only
been reached for a selected class of dimer molecules by combining the technique
of Feshbach association in ultracold, nearly quantum degenerate, atomic samples
with the technique of stimulated ground-state transfer (stimulated Raman
adiabatic passage, STIRAP) as pioneered on homonuclear Rb$_2$ and Cs$_2$
\cite{Winkler2007cot,Danzl2008qgo,Danzl2010auh} and heteronuclear fermionic KRb
\cite{Ni2008ahp}. This strategy has recently been applied to various other
heteronuclear alkali combinations, i.e., to bosonic RbCs
\cite{Takekoshi2014uds,Molony2014cou}, fermionic NaK \cite{Park2015udg}, and
bosonic NaRb \cite{Guo2016coa}. In essence, low entropy is obtained on the
atomic samples, and Feshbach association and subsequent STIRAP transfer are
aimed at maintaining low entropy.

Mixing the degenerate atomic samples is of crucial importance for our specific
purpose to create heteronuclear molecules, as well as for many other
applications \cite{Petrov2015qms,Petrov2016uld,Meinert2016boi,Gavish2005mwl,Paredes2005eqp,Roscilde2007qea}.
Maintaining low entropy in the course of the mixing process and during the
subsequent step of association to molecules poses a great experimental
challenge. Ideally, each atom from one species should find precisely one atom
from the other species for pairing up. Loss processes due to atomic three-body
recombination (before the association process) and vibrational relaxation of
Feshbach molecules as a result of either atom-molecule or molecule-molecule
collisions (after the association process) should be avoided. It is difficult
to prevent these loss processes, and all experiments with 3D bulk mixtures
\cite{Ni2008ahp,Takekoshi2014uds,Molony2014cou,Park2015udg,Guo2016coa} have
been able to convert only a comparatively small fraction of the initial
heteronuclear atomic mixture to molecules, which has led to a significant
increase of the systems' entropy.

In this Letter, we demonstrate a general sample-mixing technique that allows us
to efficiently produce heteronuclear atom pairs at the individual lattice sites
of an optical lattice and thereby to prepare low-entropy samples of
heteronuclear dimer molecules. With this technique we can mix atom samples that
are immiscible under background conditions while largely avoiding three-body
losses. Our quantum engineering approach combines superfluid transport with
interspecies interaction control and atom localization as a result of the
superfluid (SF) to Mott-insulator (MI) quantum phase transition. Specifically,
in a nontrivial generalization of the work with homonuclear molecules
\cite{Danzl2010auh}, we prepare heteronuclear Rb-Cs atom pairs at high lattice
filling by employing the SF to MI transition twice, first for the Cs sample to
create a one-atom-per-site Cs MI, then for the Rb sample on top of the Cs MI
with the aim to create a flat distribution of Rb-Cs atom pairs
\cite{Danzl2010auh}. These pairs are subsequently converted to RbCs molecules.
Control of the interspecies interaction at an interspecies Feshbach resonance's
zero crossing is needed to allow for sample mixing. Our technique minimizes
loss since the lattice greatly suppresses atomic three-body processes and
shields the Rb-Cs atom pairs and the subsequently formed RbCs molecules from
collisions. Recently, with some similarity to our work, low-entropy samples of
fermionic KRb ground-state molecules \cite{Moses2015coa,Covey2016dda} have been
produced in an optical lattice by forming atom pairs from a band insulator for K
on top of a MI for Rb.

The experiment starts with spatially separated Bose-Einstein condensates (BEC)
of $^{133}$Cs and $^{87}$Rb atoms, levitated against gravity and trapped in
crossed dipole traps as shown schematically in Figs.~\ref{Fig1}(a) and \ref{Fig1}(b)
\cite{Lercher2011poa,supmat}. The initial distance between the samples in the
horizontal plane is $x = 100$~$\mu$m. Typically, we have $4.0\times10^4$ atoms
in the Rb BEC and $1.7\times10^4$ atoms in the Cs BEC at a magnetic offset
field $B \approx 21.0~\text{G}$. At this field value, which is suitable for
producing the Cs BEC \cite{Weber2003bec}, the Rb-Rb and the Cs-Cs intraspecies
scattering lengths are $a_\text{RbRb} \approx 100$~$a_0$ and
$a_\text{CsCs} \approx 220$~$a_0$, respectively, while the interspecies
scattering length is $a_\text{RbCs} \approx 645$~$a_0$ \cite{Takekoshi2012gad},
rendering the two BECs immiscible. Here, Cs (Rb) is in the
$f_\text{Cs}\!=\!3, m_{f_\text{Cs}}\!=\!3$ ($f_\text{Rb}\!=\!1,
m_{f_\text{Rb}}\!=\!1$) spin state. A 3D cubic optical lattice, generated by
three retroreflected laser beams at $\lambda = 1064.5$~nm with large
$1/e^2$~waists of about $450$~$\mu$m covering both samples, is ramped up to
induce the SF-to-MI phase transition for the Cs sample. At a lattice depth of
$V_\text{mix}^\text{Cs} = 20$~$E_\text{rec}^\text{Cs}$, where
$E_\text{rec}^\text{Cs} = h^2/(2m_\text{Cs} \lambda^2)$ is the Cs photon recoil
energy, we create a one-atom-per-site MI for Cs \cite{supmat}. Typically, 80\%
of the atoms in the initial Cs BEC are found to be in the one-atom-Mott shell
\cite{Meinert2013qqi}. The Rb sample, which sees a depth of
$V_\text{mix}^\text{Rb} \approx 7.7$~$E_\text{rec}^\text{Rb}$, is still
superfluid. Here, $E_\text{rec}^\text{Rb} = h^2/(2m_\text{Rb} \lambda^2)$ is
the Rb photon recoil energy. The Rb sample, over the course of the next
$1500$~ms, is steered onto the Cs sample by moving the underlying dipole trap
beam linearly in time \cite{supmat}. Before the two samples start to overlap,
the field $B$ is increased to $B \approx 354.95$~G to access an interspecies
Feshbach resonance's zero crossing with slope
$d a_\text{RbCs}/dB=0.30$~$a_0/\text{mG}$ \cite{supmat}. Note that in the
vicinity of the interspecies Feshbach resonance $a_\text{CsCs}$ is
prohibitively large, $a_\text{CsCs} \approx 2500~a_0$ \cite{Berninger2013fsf}.
A mixed sample without the lattice would experience rapid Cs-Cs-Cs and Cs-Cs-Rb
three-body loss. The lattice thus assures sufficient stability over the course
of the mixing process.

%%%%%%%%%%%%%%%%%%%%%%%%%%%%%%%%%
\begin{figure}[tbp]
\includegraphics[width=\columnwidth]{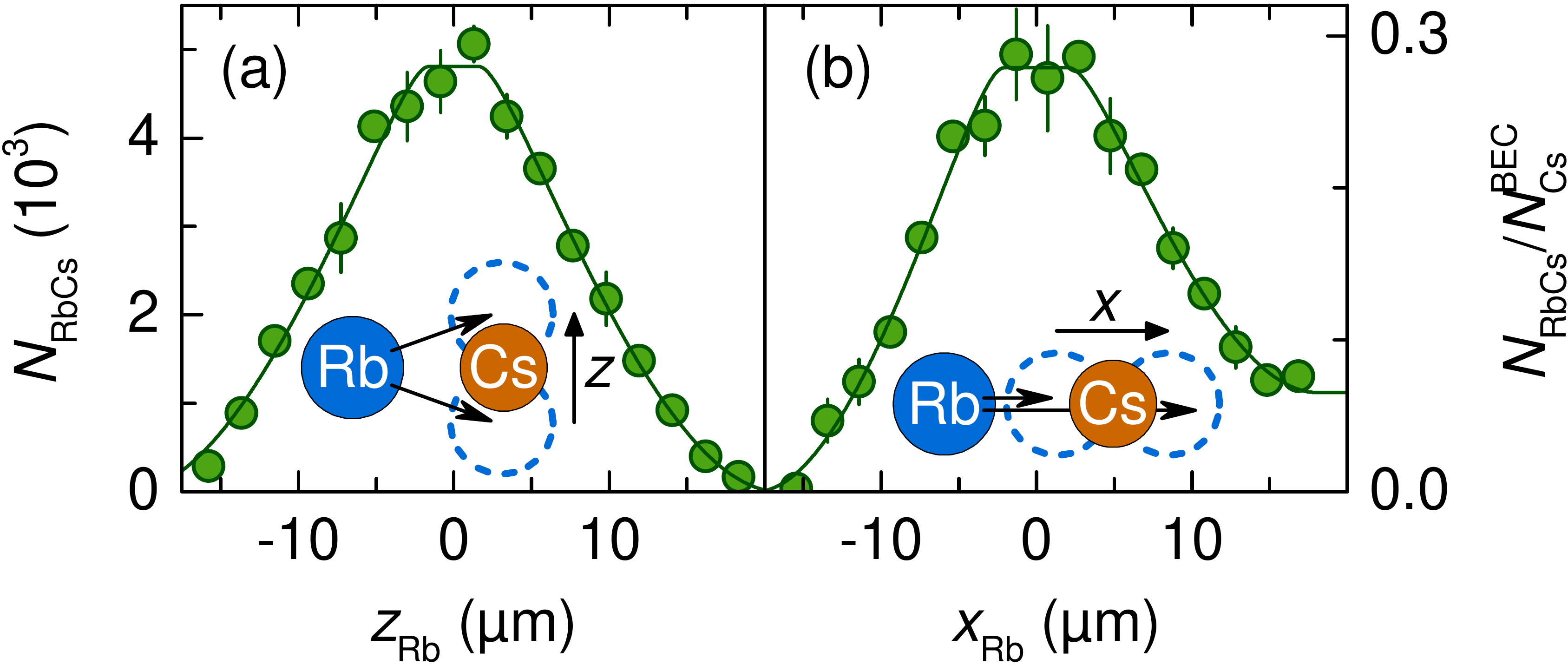}
\caption{Sample spatial overlap. Number of Feshbach molecules
$N_\text{RbCs}$ and normalized number $N_\text{RbCs}/N_\text{Cs}^\text{BEC}$ as
a function of (a)~Rb sample vertical position $z_\text{Rb}$ and (b)~Rb sample
horizontal position $x_\text{Rb}$ as indicated by the cartoons. Each data point
is the average of $3$~runs, and the error bars indicate the standard
error. The solid lines in (a) and (b) are simple fits to the data to extract
the sample radii $(r_\text{z}^\text{Cs}, r_\text{z}^\text{Rb}, r_\text{x}^\text{Cs},
r_\text{x}^\text{Rb})=(9.6(4), 11.3(2), 8.0(5), 10.1(3))$~$\mu$m, assuming the
convolution of spherical samples with homogenous densities and taking into
account the trailing tail by allowing for a one-sided offset \cite{supmat}.}
\label{Fig2}
\end{figure}
%%%%%%%%%%%%%%%%%%%%%%%%%%%%%%%%%

After spatial overlap, the lattice depth is further raised to
$V_\text{fin}^\text{Cs} = 36$~$E_\text{rec}^\text{Cs}$
($V_\text{fin}^\text{Rb} = 13.8$~$E_\text{rec}^\text{Rb}$) by a linear ramp
over the course of $40$~ms. This now induces the SF-to-MI transition on the Rb
sample on top of the Cs sample, localizing Rb atoms and, hence, producing Rb-Cs
atom pairs at the individual sites of the lattice. We note that we reach the
phase transition point but do not enter deeply into the MI regime for Rb
because of limited laser power. To quantify the success of the mixing process, we
associate the Rb-Cs atom pairs to RbCs Feshbach molecules by ramping down $B$
across the resonance pole at $352.74$~G. Further lowering $B$ to below $315$~G
allows us to enter a Feshbach state with a much lower magnetic moment than the
initial molecular state \cite{supmat}. The number of molecules $N_\text{RbCs}$
is determined via Stern-Gerlach separation of atoms and molecules in time of
flight followed by reversing the association path to dissociate the molecules
and measuring the number of atoms that have previously been bound in a molecule
\cite{Takekoshi2012gad,Takekoshi2014uds}. Note that the molecular sample is
essentially frozen in the lattice since the molecular mobility is much lower
than the atomic mobility as a result of higher mass and higher polarizability.
Each molecule resides in the lowest vibrational quantum state of its respective
lattice well \cite{Danzl2010auh}, as can be checked in a band-mapping
experiment. Our figure of merit is $N_\text{RbCs}/N_\text{Cs}^\text{BEC}$,
where $N_\text{Cs}^\text{BEC}$ is the (constant) number of Cs atoms in the
initial BEC, giving us a lower bound for the lattice filling fraction $p$
\cite{supmat}. We note that, up to this point, we have lost about 50\% of the
atoms that were in the initial Cs BEC: About 15\% of the Cs atoms are lost
irrespective of the Rb sample. An additional 35\% are lost when the two samples
are allowed to interact, presumably due to interspecies three-body
recombination. Further details on experiment timing, Feshbach structure, and
sample characterization are given in Ref.~\cite{supmat}.

First we address the samples' spatial overlap. We vary the final position of
the Rb sample in the horizonal $x$ and vertical $z$~direction and plot
$N_\text{RbCs}$, respectively, $N_\text{RbCs}/N_\text{Cs}^\text{BEC}$ as a
function of the final position as shown in Figs.~\ref{Fig2}(a) and \ref{Fig2}(b). For this
set of data, we form up to $5000$~RbCs molecules when the spatial overlap is
maximal. When varying $z$, we record data that are optimized as a function of
$x$, and vice versa. The data clearly show the spatial convolution of the two
samples. They allow us to estimate the sample extent as indicated in
Figs.~\ref{Fig2}(a) and \ref{Fig2}(b) \cite{supmat}. Also, they shows that we control the
relative positions of the two samples to much better than their sizes. When
varying the vertical position, the data are symmetric around the origin as one
may expect. However, varying the horizontal position shows a clear asymmetry.
We attribute this asymmetry to a trailing tail that the Rb sample develops
during transport. The origin of the tail and whether its existence limits our
pair formation efficiency will have to be addressed in future work.

%%%%%%%%%%%%%%%%%%%%%%%%%%%%%%%%%
\begin{figure}[tbp]
\includegraphics[width=\columnwidth]{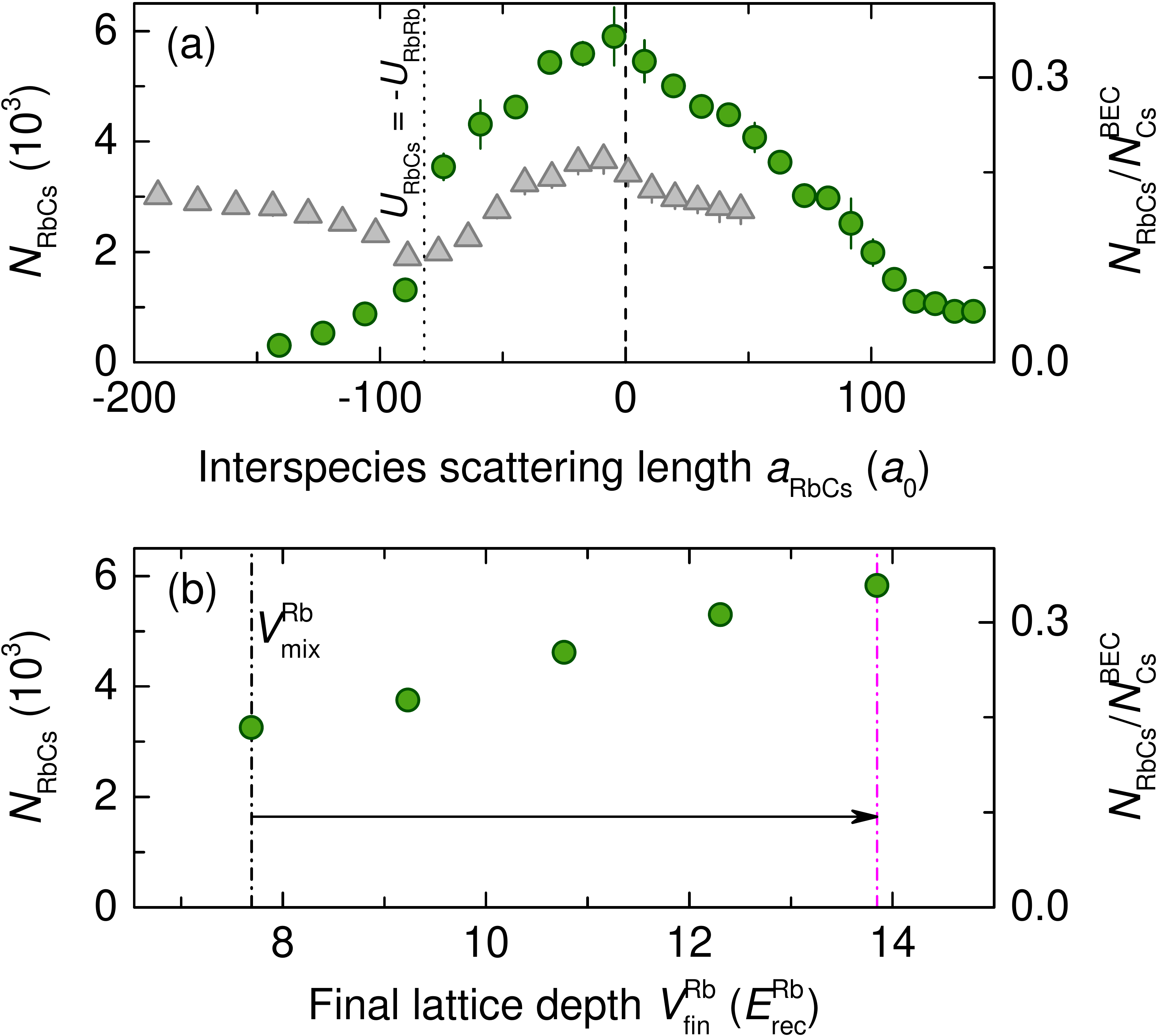}
\caption{(a)~Characterization of mixing process and subsequent
evolution. Circles: Number of Feshbach molecules $N_\text{RbCs}$, respectively,
normalized number $N_\text{RbCs}/N_\text{Cs}^\text{BEC}$ as a function of
$a_\text{RbCs}$. Triangles: $N_\text{RbCs}$, respectively,
$N_\text{RbCs}/N_\text{Cs}^\text{BEC}$ after a fixed hold time $\tau=270$~ms
upon ramping $a_\text{RbCs}$ within $10$~ms to the value as indicated. The
vertical dotted line at $a_\text{RbCs}=-82$~$a_0$ indicates the resonance
condition $U_\text{RbCs} = -U_\text{RbRb}$ as discussed in the text.
(b)~Mixing efficiency as quantified by $N_\text{RbCs}$, respectively,
$N_\text{RbCs}/N_\text{Cs}^\text{BEC}$ as a function of final lattice depth
$V_\text{fin}^\text{Rb}$ as seen by the Rb atoms. Error bars in (a) and (b)
reflect the standard error for $3$ to $11$~runs of the experiment.}
\label{Fig3}
\end{figure}
%%%%%%%%%%%%%%%%%%%%%%%%%%%%%%%%%

Next, we characterize the mixing process in terms of the interspecies
interaction strength. Figure~\ref{Fig3}(a) shows the result of our measurements
for optimal spatial overlap when we choose different values for $a_\text{RbCs}$
in the vicinity of the zero crossing during the mixing process. The efficiency
is maximal for nulled interactions and it drops off for repulsive and
attractive interactions with an initial slope of $\sim \pm 0.1/50~a_0^{-1}$. At
maximum, our figure-of-merit $N_\text{RbCs}/N_\text{Cs}^\text{BEC}$ is
significantly above 30\%. Interestingly, the data are not symmetrically
distributed about the origin. There is a steep edge at
$a_\text{RbCs} \approx -80$~$a_0$. The reason for this edge is not entirely
clear, but it coincides with the resonance condition
$U_\text{RbCs}=-U_\text{RbRb}$, where $U_\text{RbCs}$ ($U_\text{RbRb}$) is the
Hubbard onsite interaction energy \cite{supmat} for Rb-Cs (Rb-Rb) atom pairs.
When this condition is fulfilled, a Rb atom can resonantly tunnel onto a site
occupied by a Rb-Cs atom pair. This should lead to enhanced Rb-Rb-Cs three-body
loss in the course of the overlap procedure. To test this hypothesis, we have
performed an experiment in which we first prepare Rb-Cs atom pairs as
discussed before at optimal efficiency, but without the subsequent step of
Feshbach association. Instead, we hold the atom pairs for $\tau=270$~ms at
various values for $a_\text{RbCs}$. During this time, Rb atoms, although with
low mobility, are able to tunnel. We then stop the evolution and determine the
number of Rb-Cs atom pairs as before via Feshbach association, Stern-Gerlach
separation, and molecule detection. The data is also plotted in
Fig.~\ref{Fig3}(a). A clear minimum can be found for $N_\text{RbCs}$ that
coincides with the resonance condition. Note that the data show a maximum that
is slightly shifted away from $a_\text{RbCs}=0$ towards negative values. We
have no immediate explanation for this.

In the next experiment we test how the molecule production efficiency depends
on the final lattice depth $V_\text{fin}^\text{Rb}$. For this, we raise the
lattice to a value in the range between $7.7$~$E_\text{rec}^\text{Rb}$ (no
raise) and $13.8$~$E_\text{rec}^\text{Rb}$ (at full laser power available) with
a linear ramp as before, and we again determine $N_\text{RbCs}$. Note that the
Rb sample is expected to undergo the SF-to-MI phase transition at
$V_\text{crit}^\text{Rb} \approx 13.8$~$E_\text{rec}^\text{Rb}$. The data for
$N_\text{RbCs}$, shown in Fig.~\ref{Fig3}(b), show no indication of
saturation. We thus may expect that the efficiency can be improved by driving
the Rb sample more deeply into the MI phase.

\begin{figure}[tbp]
\includegraphics[width=\columnwidth]{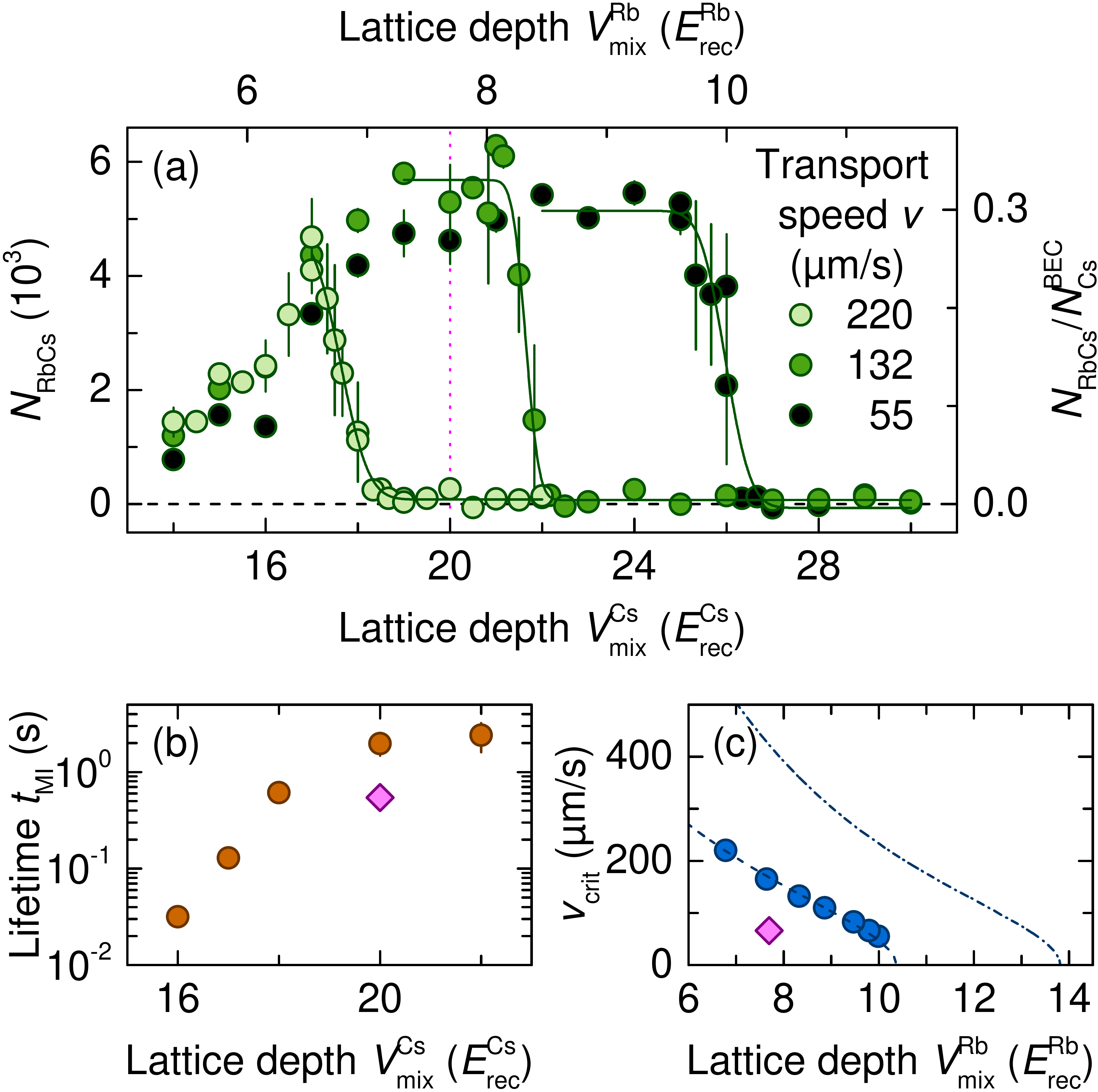}
\caption{Mixing speed limit. (a)~Mixing efficiency as
quantified by $N_\text{RbCs}$, respectively,
$N_\text{RbCs}/N_\text{Cs}^\text{BEC}$ as a function of lattice depth
$V_\text{mix}^\text{Rb}$, respectively, $V_\text{mix}^\text{Cs}$ during mixing
for various transport speeds as indicated. The solid lines are error-function
fits to determine the critical lattice depth. (b)~Lifetime $t_\text{MI}$ of the
Cs one-atom Mott insulator (circles) as a function of lattice depth
$V_\text{lat}^\text{Cs}$ at $a_\text{CsCs} \approx 2500$~$a_0$. The diamond
indicates the duration of the mixing process at the mixing lattice depth
$V_\text{mix}^\text{Cs} = 20$~$E_\text{rec}^\text{Cs}$. (c)~Critical transport
speed $v_\text{crit}$ (circles) as determined from data shown in (a) as a
function of lattice depth $V_\text{mix}^\text{Rb}$ during mixing. The diamond
indicates the conditions used in most experiments reported here. The
dashed-dotted line is the theory prediction from Ref.~\cite{Altman2005sit}. The
dashed line through the data is the same prediction but shifted horizontally by
$3.5$~$E_\text{rec}^\text{Rb}$. The error bars reflect the standard error.}
\label{Fig4}
\end{figure}

Finally, we address the question of how quickly the two samples can be merged.
Evidently, a fast merger is desirable in order to reduce the effect of loss
processes as a result of three-body recombination or heating processes due to
laser-intensity noise and beam-pointing instabilities. We vary the speed $v$ at
which we steer the Rb dipole trap towards and onto the Cs sample and then
determine $N_\text{RbCs}$ as before. The data, as a function of mixing lattice
depth $V_\text{mix}^\text{Rb}$, respectively, $V_\text{mix}^\text{Cs}$ is shown
in Fig.~\ref{Fig4}(a). For a given $v$, as $V_\text{mix}^\text{Rb}$ is
increased, the mixing process first becomes more efficient, but then
experiences an abrupt breakdown at specific values $V_\text{crit}^\text{Rb}$
that we determine by error-function fits to the data \cite{supmat}. For
sufficiently low speeds (see, e.g., the data for $v=55$~$\mu\text{m}/\text{s}$)
the efficiency saturates before it experiences the breakdown. We can explain
the reduction of efficiency towards lower $V_\text{mix}^\text{Rb}$ by a reduced
lifetime of the Cs MI state. A measurement of this lifetime at $B=355$ G,
corresponding to $a_\text{CsCs} \approx 2500$~$a_0$, is shown in
Fig.~\ref{Fig4}(b) \cite{supmat}. Below about
$V_\text{mix}^\text{Cs} \approx 20$~$E_\text{rec}^\text{Cs}$, corresponding to
$V_\text{mix}^\text{Rb} \approx 7.5$~$E_\text{rec}^\text{Rb}$, the lifetime is
significantly reduced, in agreement with the loss of efficiency below the same
value as found in the data shown in Fig.~\ref{Fig4}(a). We note that the Cs MI
lifetime is much higher at lower values for $B$, respectively, $a_\text{CsCs}$ at
which we produce the Cs BEC.

%%%%%%%%%%%%%%%%%%%%%%%%%%%%%%%%%
%%%%%%%%%%%%%%%%%%%%%%%%%%%%%%%%%

We attribute the breakdown at higher values of $V_\text{mix}^\text{Rb}$ to a
breakdown of the Rb superfluid current \cite{Altman2005sit,Mun2007pdf}. A given
depth $V_\text{mix}^\text{Rb}$ thus defines a critical transport speed
$v_\text{crit}$. Figure~\ref{Fig4}(c) plots this critical speed as determined
from data shown in Fig.~\ref{Fig4}(a) as a function of
$V_\text{mix}^\text{Rb}$. Higher values of $V_\text{mix}^\text{Rb}$ determine a
lower $v_\text{crit}$. Simple linear extrapolation of our data towards larger
values of $V_\text{mix}^\text{Rb}$ gives a zero critical speed at
$V_\text{mix}^\text{Rb} \approx 11$~$E_\text{rec}^\text{Rb}$.
Figure~\ref{Fig4}(c) also plots the theory prediction of
Ref.~\cite{Altman2005sit}. Evidently, our data is significantly below this
prediction. Note that shifting the theory curve horizontally by
$3.5$~$E_\text{rec}^\text{Rb}$ gives very good agreement. We have no
explanation for the discrepancy between our data and the theory prediction. We
note that for the data shown, e.g., in Fig.~\ref{Fig3}(a) we have chosen a
transport speed that is significantly below the critical speed, as indicted by
the diamond in Fig.~\ref{Fig4}(c).

Evidently, our mixing procedure can be improved in various ways. Apart from
going more deeply into the Rb MI phase one should increase the speed for
superfluid transport and reduce the initial distance between the two samples to
minimize the time over which three-body loss processes take place. Ideally,
one should place two pancake-shaped BECs directly parallel to each other and
then merge the two samples along the tightly confined direction.

In conclusion, we have developed a procedure to efficiently mix two-species
bosonic samples and have applied it to form low-entropy samples of bosonic RbCs
Feshbach molecules in an optical lattice. The entropy per molecule can be
estimated to $\approx 2$~$k_\text{B}$ \cite{supmat}. Note that this is an upper
bound, as we, at this point, can only give a lower bound on the lattice filling
fraction. Future experiments with interacting dipoles will allow us a more
precise determination of the filling fraction \cite{Hazzard2014mbd}. Our mixed
bosonic two-species sample is an ideal starting point for experiments in the
context of quantum many-body physics, e.g. on strongly interacting two-species
Bose polarons \cite{Hu2016bpi}, on quantum droplet formation in Bose mixtures
\cite{Petrov2015qms,Petrov2016uld}, on impurity transport
\cite{Meinert2016boi}, and on disorder in Bose-Bose mixtures
\cite{Gavish2005mwl,Paredes2005eqp,Roscilde2007qea}. Low-entropy dipolar
quantum gases can be generated by transferring the molecules via STIRAP to the
electronic and rovibrational ground state. We expect STIRAP transfer
efficiencies exceeding 90\% as demonstrated in a previous publication
\cite{Takekoshi2014uds}. Similar rovibrational wave-function overlap but better
sample localization will allow us to push efficiencies towards 100\%. When in
the ground state and polarized by an external electric field, the dynamics of
the gas will be dominated by nearest-neighbor interactions with interaction
strength on the order of $h \times 1$~kHz. This will allow us to study
important problems in quantum many-body physics, such as the phase diagram of
the Bose-Hubbard model extended by a long-range interaction term
\cite{Damski2003coa,Capogrosso2010qpo} and quantum spin models with long-range
interactions \cite{Hazzard2014mbd}.

We thank S.~Mezinska for technical assistance and M.~J.~Mark and F.~Meinert for
discussions. We are grateful to C.~R.~Le~Sueur for providing detailed
data for $a_\text{RbCs}$ and on the molecular structure as
show in Fig.~S2 of the Supplemental Material. We acknowledge support by the
Austrian Science Fund (FWF) through the Spezialforschungsbereich (SFB) FoQuS
within project P06 (FWF Project No.\ F4006-N23).

%\bibliography{ultracold_molecules_HCN,ultracold}

\clearpage

\setcounter{figure}{0}
\renewcommand\thefigure{S\arabic{figure}}

\section{Supplemental Material}

\subsection{Experimental timing}
\begin{figure}[tbp]
\includegraphics[width=\columnwidth]{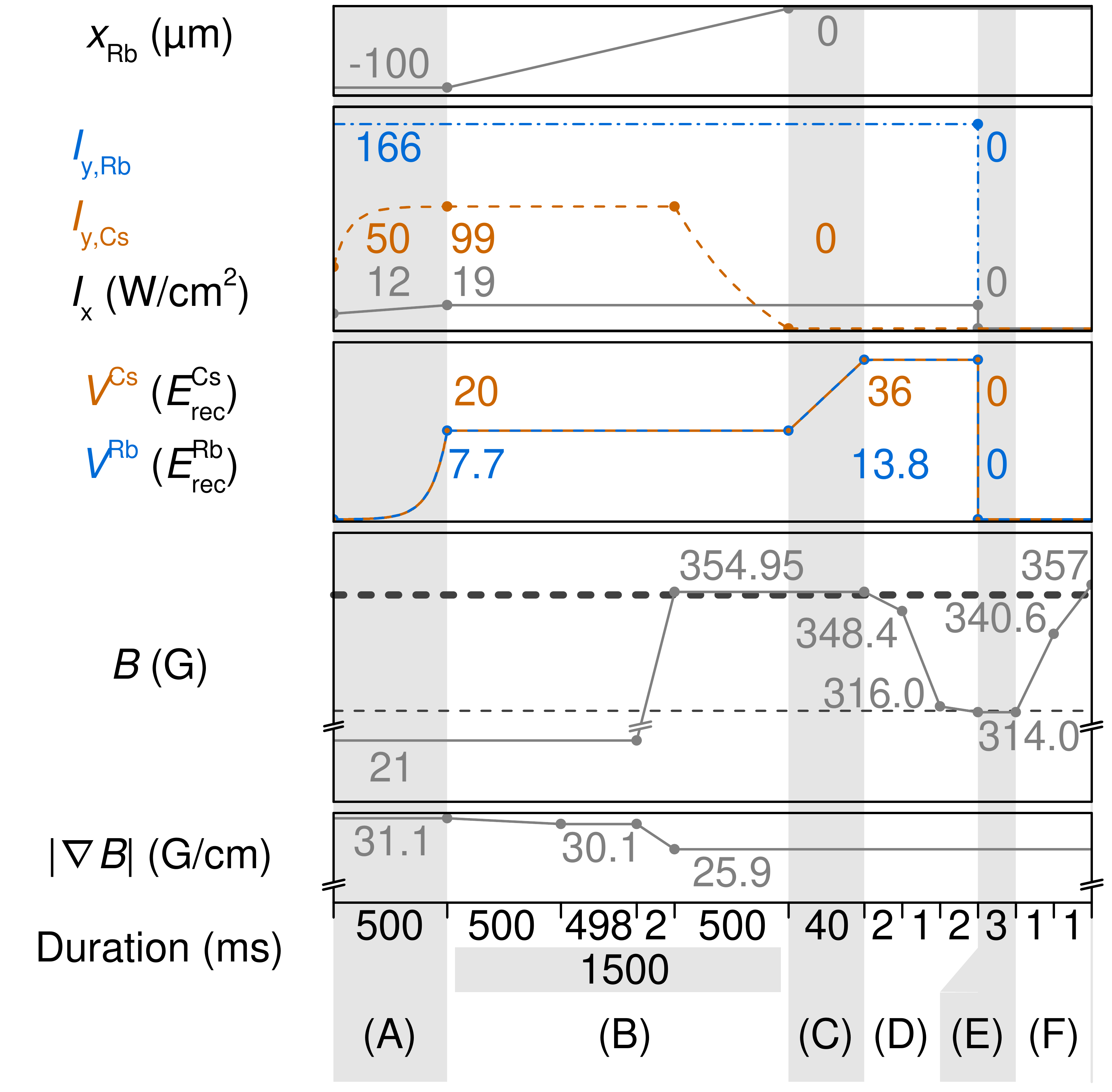}
\caption{Typical timing sequence for the experimental cycle
with typical values for the Rb trap position $x_\text{Rb}$ along the transport
direction, the beam intensities $I_\text{y,Rb}$, $I_\text{y,Cs}$, and
$I_\text{x}$, the lattice depth $V^\text{Cs} = 2.6 V^\text{Rb}$, the magnetic
offset field $B$, and the magnetic field gradient $\left| \nabla B \right|$ as
a function of time. The various experimental stages (A) lattice loading, (B) Rb
transport, (C) Rb localization, (D) molecule formation, (E) Stern-Gerlach
separation, and (F) dissociation and detection are indicated in the lower part
of the diagram. The horizontal axis is not to scale.}
\label{Reichsoellner_supmat_Fig1}
\end{figure}

Details of a typical experimental sequence are given in
Fig.~\ref{Reichsoellner_supmat_Fig1}. The sequence starts with the loading of
the two BECs into the optical lattice and creating a Cs MI, followed by the
transport of the Rb BEC, the Rb-Cs pair formation, the production and the
subsequent detection of molecules. We denote the intensity in the dipole
trapping beams by $I_\text{y,Rb}$, $I_\text{y,Cs}$, and $I_\text{x}$ for the
beams along the $y$~direction for the inital Rb and Cs dipole traps and for the
beam joining the two samples along the transport $x$~direction, respectively
(see Fig.~1(a) of the main article). In the course of the sequence, the lattice
depth $V^\text{Cs} = 2.6 V^\text{Rb}$, the Rb trap position $x_\text{Rb}$ along
the transport direction, the intensities $I_\text{y,Rb}$, $I_\text{y,Cs}$, and
$I_\text{x}$, the magnetic offset field $B$, and the magnetic field gradient
$\left| \nabla B \right|$ are ramped. Note that the Cs atoms see a $1.08$ times
larger trap frequency than the Rb atoms for a given dipole trap laser power.

Initially, the Rb BEC and the Cs BEC are spatially separated along the
$x$ direction by $\approx 100$~$\mu$m with $\nu_{\text{x,Rb}}=38$~Hz,
$\nu_{\text{x,Cs}} = 10$~Hz, $\nu_{\text{y,Rb}}=14$~Hz,
$\nu_{\text{y,Cs}}=15$~Hz, $\nu_{\text{z,Rb}} = 39$~Hz, and
$\nu_{\text{z,Cs}} = 18$~Hz. Here, $\nu_{\text{x,Rb}}$ denotes the trap
frequency for the Rb trap in $x$~direction, and analogously for the other trap
frequencies. The initial distance of the two BECs is large enough to avoid
spilling from one sample into the other. In the course of the transport, while
the confinement of the two samples is controlled by the intensities
$I_\text{y,Rb}$, $I_\text{y,Cs}$, and $I_\text{x}$, it is strongly modified
along the $x$~direction when the underlying dipole traps start to overlap. Beam
steering to move the Rb trap along the $x$ and the $z$~directions is achieved
by a two-axes translational piezo flexure stage onto which we have attached the
fiber tip of the fiber that delivers the light for the Rb trap beam propagating
in the $y$~direction. Beam steering is done with $\mu$m precision as verified
by in-situ absorption images. The initial gradient
$\left| \nabla B \right| = 31.1$~$\text{G}/\text{cm}$ levitates the Cs sample
against gravity, but slightly overlevitates the Rb sample.

Upon loading both ensembles into the 3D optical lattice with a lattice spacing
of $\lambda/2=532.25$~nm we drive the SF-to-MI phase transition for the Cs
sample while leaving the Rb sample superfluid. For this, we exponentially
increase the lattice depth to
$V^\text{Cs} = V_\text{mix}^\text{Cs} = 20$~$E_\text{rec}^\text{Cs}$
($V_\text{mix}^\text{Rb} = 7.7$~$E_\text{rec}^\text{Rb}$). At the same the time
underlying Cs dipole trap is stiffened to assure that we create a clean
one-atom MI shell for Cs.

Subsequently, the Rb sample is transported through the lattice towards and onto
the Cs sample within typically $1500$~ms by moving the underlying Rb dipole
trap along the $x$~direction. The gradient $\left| \nabla B \right|$ is ramped
to $30.1$~$\text{G}/\text{cm}$ in the course of the first $500$~ms of the
transport process for optimum levitation of Rb. We slightly adjust the vertical
position of the Rb trap in the course of the transport for optimal overlap with
the Cs sample.

After $998$~ms of the transport, i.e. $502$~ms before the Rb sample has reached
its final position, the offset field $B$ is ramped within $2$~ms to the zero
crossing for $a_\text{RbCs}$ at $354.95$~G. Simultaneously
$\left| \nabla B \right|$ is adjusted to $25.9$~$\text{G}/\text{cm}$ to
compensate for the change in the magnetic moment of Rb as $B$ is changed.
Within the last $500$~ms of the transport the underlying Cs dipole trap beam
along the $y$~direction is adiabatically turned off to avoid its influence on
the final phase of the Rb transport.

%%%%%%%%%%%%%%%%%%%%%%%%%%%%%%%%%
%%%%%%%%%%%%%%%%%%%%%%%%%%%%%%%%%

\begin{figure}[tbp]
\includegraphics[width=\columnwidth]{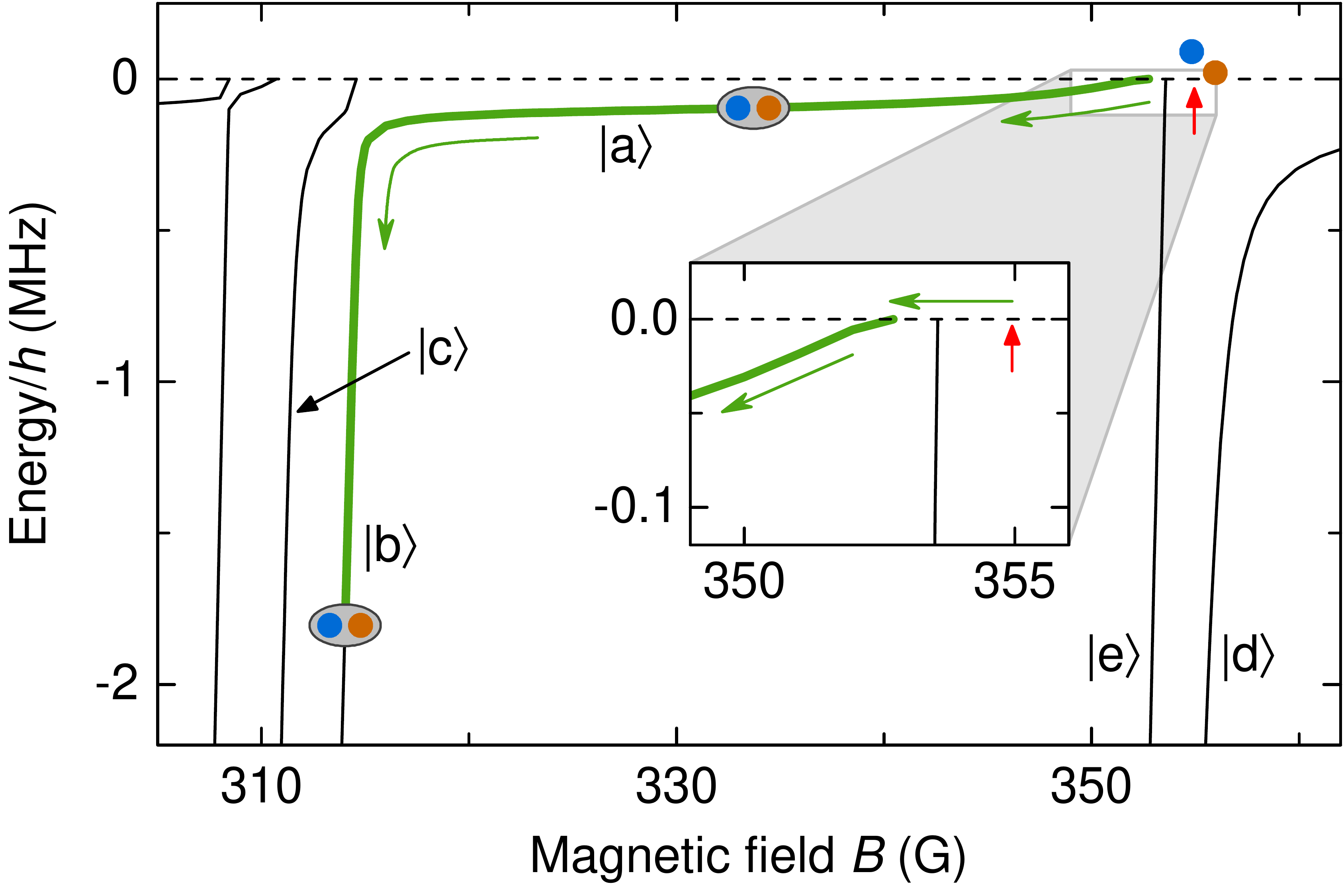}
\caption{Zeeman diagram for the molecular states with
$M_\text{tot}=4$ just below the ground-state two-atom
$(f_\text{Rb},f_\text{Cs})=(1,3)$ threshold. Energies as a function of $B$ are
given relative to the field-dependent dissociation threshold (horizonal dashed
line). The red arrow marks the position of the Feshbach-resonance zero crossing
at which Rb-Cs atom pairs are produced. The magneto-association path is
indicated by the green arrows and the thick green line, first following state
$|a\rangle$ and then ending in state $|b\rangle$ as $B$ is lowered. The inset
zooms into the resonance region, showing the zero-crossing position at
$354.95$~G (red arrow), an overlapping molecular state $|e\rangle$ hitting
threshold at $353.57$~G (black, nearly vertical line) and the pole of the
Feshbach resonance at $352.74$~G where state $|a\rangle$ hits threshold.}
\label{Reichsoellner_supmat_Fig2}
\end{figure}

As soon as the Rb transport is finished we increase the lattice depth
$V^\text{Cs}$ to $V_\text{fin}^\text{Cs} = 36$~$E_\text{rec}^\text{Cs}$
($V_\text{fin}^\text{Rb}=13.8$~$E_\text{rec}^\text{Rb}$) to drive the Rb sample
into the MI regime and to form Rb-Cs atom pairs in the lattice. By
adiabatically sweeping $B$ within $2$~ms over the pole of the Feshbach
resonance at $355.74$~G to $348.4$~G we associate the paired atoms to weakly
bound molecules (see the Zeeman diagram in
Fig.~\ref{Reichsoellner_supmat_Fig2}). Subsequently, within $1$~ms, we jump $B$
to $316.0$~G, where an avoided crossing is located that allows us to transfer
the molecules into the second state by ramping $B$ in $2$~ms to $314.0$~G (see
Fig.~\ref{Reichsoellner_supmat_Fig2} and the discussion in the next section).
This final molecular state possesses a magnetic moment that significantly
differs from the one of the unbound atoms, allowing Stern-Gerlach separation of
molecules and atoms. We switch off the dipole trap and lattice potentials
abruptly and spatially separate the molecules from nonassociated atoms within
$3$~ms time of flight. For detection we reverse the Feshbach association ramp
to dissociate the molecules back to atoms. These atoms and also the atoms that
were not subject to molecule formation are subsequently detected by standard
absorption imaging.

%%%%%%%%%%%%%%%%%%%%%%%%%%%%%%%%%
%%%%%%%%%%%%%%%%%%%%%%%%%%%%%%%%%

\subsection{Feshbach states}
In the following we adopt the labelling
$|n(f_\text{Rb},f_\text{Cs})L(m_{f_\text{Rb}},m_{f_\text{Cs}})M_\text{tot}\rangle$
of the Feshbach states used in Ref.~\cite{Takekoshi2012gad2}, where a broader
overview on the Rb-Cs Feshbach states and resonances is given. The number $n$
indicates the vibrational level as counted from threshold (i.e.\ $n=-1$ refers
to the least bound vibrational level, $n=-2$ is the second least bound
vibrational level, etc.)\ and the quantum numbers $f$ and $m_f$ label the total
atomic angular momentum and its projection onto the $B$-field axis,
respectively. The quantum number $L$ (with $s$ for $L=0$, $p$ for $L=1$, $d$
for $L=2$ etc.)\ denotes the molecular rotational angular momentum and
$M_\text{tot}=m_{f_\text{Rb}}+m_{f_\text{Cs}}+M_L$ is the sum of all projected
angular momenta. Since here $M_\text{tot}=4$ for all relevant states we omit
$M_\text{tot}$.

For the mixing process and molecule creation we make use of a $2.21$-G-wide
Rb-Cs Feshbach resonance, whose pole is located at $352.74$~G. The background
interspecies scattering length in the vicinity of this resonance is
$a_\text{RbCs}=654$~$a_0$ \cite{Ruth2}. The Rb and Cs samples are initially
overlapped at the zero crossing for $a_\text{RbCs}$ at $354.95$~G, indicated by
the red arrow in Fig.~\ref{Reichsoellner_supmat_Fig2}. From there $B$ is ramped
to the pole of the resonance, jumping nonadiabatically over a very narrow Feshbach
resonance (with a width of $1.2$~mG) located at $353.57$~G caused by the
molecular state $|e\rangle=|-6(2,4)d(0,2)\rangle$ \cite{Ruth2} (see inset in
Fig.~\ref{Reichsoellner_supmat_Fig2}). The atoms then enter a Feshbach state
that initially is a combination of state $|a\rangle=|-1(2,4)s(1,3)\rangle$ and
$|d\rangle=|-6(2,4)s(0,4)\rangle$ (green curve in the inset to
Fig.~\ref{Reichsoellner_supmat_Fig2}). Upon further lowering $B$ the state
becomes $|a\rangle$. In order to allow the separation of molecules from unbound
atoms via the Stern-Gerlach technique, an anti-crossing of molecular states at
$315$~G is used to adiabatically transfer the molecules from state $|a\rangle$
to $|b\rangle$. The latter one is together with state $|c\rangle$ a
superposition of the molecular states $|-6(2,4)d(0,3)\rangle$ and
$|-6(2,4)s(1,3)\rangle$. The exact composition of the molecular states is of
importance for future work when the molecules are to be transferred by STIRAP
from their weakly-bound state to their rovibronic ground state.

\subsection{Convolution modeling}

We characterize the overlap of the two atomic samples, as presented in Fig.~2
of the main article, by using a simple model that assumes two spheres of
homogeneous density with radii $r_1$ and $r_2$. The convolution $C(d)$ of two
spheres as a function of the distance $d$ between their centers is given by
\begin{equation*}
  C(d) = \frac{\pi}{12d}(r_1+r_2-d)^2 \left[ d^2 + 2d(r_1+r_2) -3(r_1-r_2)^2 \right]
\end{equation*}
for partial overlap $\left| r_1 - r_2 \right| < d < (r_1+r_2)$. We allow for an
asymmetric offset $N_\text{RbCs}^\text{offset}$ to account for the trailing
tail that is observed in the experiment when convolving the samples along the
transport $x$~direction. Setting $x_\text{Cs}=0$ for the center of the Cs
sample, the full fit function to the number of RbCs molecules $N_\text{RbCs}$
then reads
\begin{widetext}
\begin{equation*}
	N_\text{RbCs}(x_\text{Rb}) = \left\{\begin{array}{lrcccr}
			0, &&&x_\text{Rb}&\leq&-(r_1+r_2) \\ N_\text{RbCs}^\text{max}\frac{C(-x_\text{Rb})}{C\left(\left|r_1-r_2\right|\right)} & -(r_1+r_2)&<&x_\text{Rb}&<&-\left|r_1-r_2\right| \\
			N_\text{RbCs}^\text{max}, & -\left|r_1-r_2\right|&\leq&x_\text{Rb}&\leq&\left|r_1-r_2\right| \\ (N_\text{RbCs}^\text{max}-N_\text{RbCs}^\text{offset})\frac{C(x_\text{Rb})}{C\left(\left|r_1-r_2\right|\right)} + N_\text{RbCs}^\text{offset}, & \left|r_1-r_2\right|&<&x_\text{Rb}&<&(r_1+r_2) \\
			N_\text{RbCs}^\text{offset}, & (r_1+r_2)&\leq&x_\text{Rb}&& \\
		\end{array}\right.
\label{equation2015112701}
\end{equation*}
\end{widetext}
For $r_1 \neq r_2$ the function has a characteristic volcanolike profile,
where $N_\text{RbCs}^\text{max}$ defines the height the plateau when one sphere
is fully enclosed by the other. The fit cannot attribute $r_1$ and $r_2$ to the
individual species, however for a sufficiently large difference in the atom
number we presume that the smaller radius belongs to the smaller sample (here
Cs). Even though the atom samples possess some ellipticity due to the shape of
their trapping potential and even though their densities are inhomogeneous, the
fit should yield acceptable estimates for samples' radii along the direction of
the convolution measurement. When compared to a simple calculation that assumes
the formation of defect-free single-shell MIs the sample radii that result from
the convolution-measurement fits are roughly $1$~$\mu$m larger. This deviation
is probably the result of a reduced density in the outer region of the MI as a
result of nonzero temperatures. Note that the convolution measurement shows
that we control the relative positioning of the two atom samples along the two
steering axes to a much higher precision than the resolution of our imaging
setup. It in particular allows us to calibrate a roughly $5$-$\mu$m chromatic
offset between the absorption images of the two species.

\subsection{Filling fraction and entropy}

We normalize the number of RbCs molecules $N_\text{RbCs}$ by the number
$N_\text{Cs}^\text{BEC}$ of atoms in the Cs BEC and thereby obtain a lower
bound for the filling fraction of the RbCs molecules in the optical lattice
when we assume a filling fraction of unity for the Cs MI right after lattice
loading and full coverage of the Cs sample by a homogeneous density of Rb
atoms. The first assumption is well fulfilled for the center region of the Cs
MI (see section Cs MI characterization). The second assumption, however, is not
necessarily fulfilled and can locally lead to an underestimation of the filling
fraction. The Rb sample is elongated along the $y$~axis, therefore a
significantly higher Rb atom number is required to fully cover the nearly
spherical Cs MI. Although the initial Rb BEC is about twice as large as the Cs
BEC, this requirement is not necessarily met, since $20$\%--$50$\% of the Rb
atoms are stuck in the lattice during transport. Furthermore, the Rb density is
not homogeneous, especially since we are not yet able to drive the Rb sample
deep into the MI regime. We note that the elongation of the Rb sample is
actually necessary to provide sufficient overlap since we have no handle on the
relative position of the atom clouds along the $y$~axis and a small offset
exists due to the fact that the focus of the Cs trapping beam is axially shifted
from the intersection with the trapping beam that defines the transport axis.

Several aspects limit the filling fraction of the RbCs molecules: The finite
lifetime of the Cs MI at high magnetic fields (see section Cs MI
characterization) as well as losses during the overlapping process, which we
attribute to interspecies three-body recombination, create vacancies in the Cs
MI. Finite coverage of the Cs MI by the Rb sample and variations in the Rb
density from $n_\text{Rb}=1$ limit the overall pair formation efficiency. The
loss of Cs atoms in absence of the Rb sample is about $15$\%. The number of Rb
atoms that are lost during the mixing process (excluding atoms that are stuck
on the way to the Cs sample during the transport) is about $1.6(1)$ times the
number of additionally lost Cs atoms in presence of Rb. It therefore appears
that this loss is dominated by the Rb-Rb-Cs three-body loss mechanism, further
reducing the number of Cs atoms by $35$\% of the Cs BEC size. We observe an
enhancement of the interspecies loss for slower transport velocities, which can
be compensated by bringing the lattice depth $V_\text{mix}$ closer to the
critical value at which the superflow of Rb breaks down. This is not surprising
considering that both the critical transport velocity and the three-body loss
rate scale with the tunneling time in the optical lattice. In the end, about
$60$\% of the remaining Cs atoms ($30$\% of $N_\text{Cs}^\text{BEC}$) are
paired with Rb and successfully form RbCs molecules.

For calculation of the molecules' entropy we assume that lattice sites occupied
by unassociated atoms can be emptied, e.g. by a blow-away technique
\cite{Thalhammer2006llf2,Danzl2010auh2}, without affecting the molecular sample.
The probability for an empty site is then $1-p$, where $p$ is the molecular
filling fraction, and the entropy per molecule is \cite{Budker2008apa2}
\begin{equation*}
  s \approx \frac{k_\text{B}}{p} \left( p \ln \left( p \right) + \left( 1-p \right) \ln \left( 1-p \right) \right).
\end{equation*}
Taking $p=30\%$ as a lower bound gives $ s = 2$~$k_\text{B}$. This value compares
well with the one reported in Ref.~\cite{Moses2015coa2}.

\subsection{On-site interspecies interaction}

For moderate values of the interspecies scattering length $a_\text{AB}$ the
Hubbard on-site interaction between atoms A and B is given by
\begin{equation*}
  U_\text{AB}=\frac{4 \pi \hbar^2}{2\mu_\text{AB}} a_\text{AB} \int{w_\text{A}^{*}(\bm{r}) w_\text{B}^{*}(\bm{r}) w_\text{B}(\bm{r}) w_\text{A}(\bm{r}) d^3r.}
\end{equation*}
Here, $\mu_\text{AB} = m_\text{A}m_\text{B}/(m_\text{A}+m_\text{B})$ is the
reduced mass and $w_{\text{A},\text{B}}$ are the Wannier functions for A and B,
respectively. Independent of the depth of the lattice we find that
$U_\text{RbCs} = - U_\text{RbRb}$ for $a_\text{RbCs} = -82$~$a_0$.

\subsection{Cs MI characterization}

At low values for $B$ (i.e. $B=21$~G, where we form the Cs BEC) the Cs sample
in the optical lattice is stable for many seconds irrespective of the lattice
depth. Increasing $B$ to high values (necessary to access the Rb-Cs Feshbach
resonance and to tune the interspecies interaction) however changes the Cs
intraspecies scattering length $a_\text{CsCs}$ dramatically. For $B=354.95$~G,
where we perform the mixing procedure, $a_\text{CsCs}$ reaches about
$2500$~$a_0$ \cite{Berninger2013fsf2}. We emphasize that in this regime it is of
crucial importance to drive the system deeply into the MI state to protect the
Cs MI from enhanced Cs-Cs-Cs three-body loss.

We analyze the stability of the Cs MI at high values for $B$ (see Fig.~4(b) of
the main article) by preparing first a Cs MI at $B=21$~G for a lattice depth
$V^\text{Cs}$. We increase $B$ within $2$~ms to $B=354.95$~G and then hold the
sample for a variable hold time $\tau$ before lowering $B$ again back to
$B=21$~G within another $2$~ms. In the end the lattice is ramped down
adiabatically and the BEC fraction of the Cs sample is determined. We find that
the BEC fraction decays exponentially with $\tau$ and determine the
$1/e$ lifetime $t_\text{MI}$ as a function of $V^\text{Cs}$. The result is
plotted in Fig.~4(b) of the main article. We note that the decay of the Cs atom
number due to Cs-Cs-Cs three-body losses is in general slower than the decay of
the BEC fraction. To maintain a stable Cs MI state over the course of the
$540$~ms that the atoms spend at high values for $B$ during the mixing process
(see section Experimental timing) the mixing lattice depth
$V_\text{mix}^\text{Cs}$ has to be at least $20$~$E_\text{rec}^\text{Cs}$
(diamond in Fig.~4(b)). Below that value a significant reduction in the RbCs
molecule production efficiency is observed (see Fig.~4(a)).

We check that we produce a relatively clean single-atom MI shell by using a
technique described in Ref.~\cite{Meinert2013qqi2}. The lattice depth along the
vertical axis is lowered to $10$~$E_\text{rec}^\text{Cs}$ to allow the atoms to
tunnel and a potential tilt along this axis is applied adiabatically. When the
tilt per lattice site approaches the Cs on-site interaction $U_\text{CsCs}$ a
quantum phase transition is driven to a density-wave ordered state where every
other lattice site along the tilt axis is occupied by a Cs-Cs atom pair. The Cs
pairs are associated to weakly bound Cs$_2$ molecules by means of a magnetic Cs
Feshbach resonance to determine the number of Cs pairs that were formed this
way. We find that about $80$\% of the entire Cs sample forms pairs, comparable
to the results presented in Ref.~\cite{Meinert2013qqi2}. Such a high pair
formation efficiency can only be achieved in a low-defect MI shell.

\subsection{Critical transport velocity}

The lattice depth $V_\text{mix}$ that is used for mixing the two samples sets a
critical transport velocity $v_\text{crit}$ for transport of Rb
\cite{Altman2005sit2}. Beyond $v_\text{crit}$ the superflow becomes chaotic in
regions with average filling $n_\text{Rb} = 1$, triggering superflow breakdown
for the entire Rb sample. The Rb sample gets stuck close to its initial
position and cannot follow the movement of the underlying Rb dipole trap.

In the vicinity of the SF-to-MI transition, for a 3D lattice, the critical
quasimomentum that corresponds to the critical transport velocity
$v_\text{crit}$ is given by \cite{Altman2005sit2}
\begin{equation*}
  q_\text{crit} = \frac{{\hbar}k_\text{rec}}{\pi} \text{Re}\!\left[ \sqrt{2(1-u/u_\text{c})} \right].
\end{equation*}
Here, ${\hbar}k_\text{rec} = h/ \lambda$ is the recoil momentum set by the
lattice light, and the interaction strength is $u = U_\text{RbRb}/J_\text{Rb}$,
with the Hubbard interaction parameter $U_\text{RbRb}$ and the tunneling matrix
element $J_\text{Rb}$ for Rb. The critical value for $u$, which marks the
SF-to-MI quantum-phase transition \cite{Jaksch1998cba2}, is
$u_\text{c} \approx 34.8$, which is reached for
$V_\text{mix}^\text{Rb} \approx 13.8$~$E_\text{rec}^\text{Rb}$. The group
velocity with which the Rb sample moves through the lattice is defined by the
dispersion relation of the lowest lattice band $E(q)$ as
$v_\text{g}(q) = \partial E(q)/\partial q$ and allows us to calculate the
critical transport velocity $v_\text{crit} = v_\text{g}(q_\text{crit})$
(dash-dotted line in Fig.~4(c) of the main article).

The value for the critical quasimomentum $q_\text{crit}$ predicted in
Ref.~\cite{Altman2005sit2} was experimentally tested in Ref.~\cite{Mun2007pdf2}
with a Rb sample in a sinusoidally moving 3D optical lattice and good agreement
between theory and experiment was found. In our experiment we perform linear
transport by means of the underlying dipole trap that moves at constant speed
in the presence of a stationary 3D lattice. The breakdown in superfluid
transport is reflected by an abrupt loss of pair formation efficiency as shown
in Fig.~4(a) of the main article. Concomitantly, we find from in-situ
absorption images that the Rb sample experiences sudden inhibition of
transport. The critical transport velocity $v_\text{crit}$ as determined from
our data is much lower by at least a factor $2$ than the one predicted by
theory. Interestingly, the data are fit well by shifting the theoretical values
by $3.5$~$E_\text{rec}^\text{Rb}$ towards shallower lattice depths (dashed line
in Fig.~4(c)). The discrepancy between our experiment and theory merits further
investigations.

\subsection{Magnetic field calibration}

The magnetic field $B$ is calibrated by microwave spectroscopy. To assure that
we work under the same conditions as in the molecule formation experiment we
use the sequence for molecule creation in an optical lattice as described above
(see also Fig.~\ref{Reichsoellner_supmat_Fig1}) but load only a Cs sample,
create a Cs one-atom MI, and substitute the molecule creation procedure by a
$20$-ms microwave pulse. A microwave antenna, powered by a programmable
microwave source with $5$~W output, irradiates the Cs atoms and drives the
$\pi$ transition of the lowest Cs hyperfine states
$|f_\text{Cs},m_{f_\text{Cs}}\rangle=|3,3\rangle$ to $|4,3\rangle$. We measure
the number of Cs atoms in state $|3,3\rangle$ and $|4,3\rangle$ by means of
Stern-Gerlach separation as we scan the microwave frequency. We extract the
resonance position from Gaussian fits to the data and calculate the
corresponding value for $B$ via the Breit-Rabi formula \cite{Steck2003cdl2}. The
values for $B$ can then be converted to values for $a_\text{RbCs}$ \cite{Ruth2}.
In this way we are able to determine the zero-crossing of the interspecies
scattering length $a_\text{RbCs}$ at $354.95$~G with an accuracy of $17$~mG. We
note that the pole of the interspecies Feshbach resonance together with its
adjacent zero crossing leave significant signatures in a different set of
experiments in which we probe the interference contrast of the superfluid Rb
sample as we scan across the Feshbach resonance. In particular, at a certain
value for $B$ we observe a drastic loss of interference contrast. This we
associate with the pole of the resonance. The position of the pole agrees with
what we expect from our $B$-field calibration.

Magnetic field stability is crucial during the mixing process when
$B \approx 355$~G. The most dominant magnetic noise contributions are at
multiples of the line frequency $50$~Hz. A feed-forward technique allows us to
suppress Fourier components at $50$~Hz, $100$~Hz, $150$~Hz, and $250$~Hz. For
this, a pick-up coil senses their phase and amplitude. During the mixing
procedure these Fourier components are then added to $B$ with a $\pi$ phase
shift synchronized to line. We thereby suppress the total magnetic field noise
down to $50$~mG rms, corresponding to about $15$-$a_0$ rms uncertainty for the
interspecies scattering length in the vicinity of its zero crossing. The
magnetic field gradient $\left| \nabla B \right| =25.9$~$\text{G}/\text{cm}$
causes an additional shift of $16$~$a_0$ across the typical $18$-$\mu$m
atom-sample diameter.

%\bibliography{ultracold_molecules_HCN,ultracold}

\end{document}